\documentclass[aip,graphicx]{revtex4-1}
\usepackage{graphicx}

\draft 

\begin{document}


\title{Anomalous magnetisation  in a $Fe_3Pt-FeRh$ film occurring at T$\approx$120 K in the field-cooled-cooling curves for low magnetic fields.}
\author{S. Salem-Sugui Jr.}
\affiliation{Instituto de Fisica, Universidade Federal do Rio de Janeiro,
21941-972 Rio de Janeiro, RJ, Brazil}
\author{A. D. Alvarenga}
\affiliation{Instituto Nacional de Metrologia Normaliza\c{c}\~ao e
Qualidade Industrial, 25250-020 Duque de Caxias, RJ, Brazil.}
\author{R.D. Noce}
\affiliation{Instituto de Qu'mica, UNESP, 1801-400, Araraquara, SP, Brazil.} 
\author{C. Salazar Mejia}
\affiliation{Instituto de Fisica, Universidade Federal do Rio de Janeiro,
21941-972 Rio de Janeiro, RJ, Brazil}
\author{H. Salim}
\affiliation{Instituto de Fisica, Universidade Federal do Rio de Janeiro,
21941-972 Rio de Janeiro, RJ, Brazil}
\author{F.G. Gandra}
\affiliation{Instituto de Fisica, UNICAMP, Campinas, SP, Brazil}
\date{\today}
\begin{abstract}
We report on an anomalous magnetization observed with temperature for low magnetic fields applied in the plane of a film formed by a thin layer of Fe-Rh deposited on a thin foil of ordered Fe$_3$Pt. The anomalous effect resembles a metamagnetic transition and occur only in the field-cooled-cooling magnetization curve at temperatures near 120 K. We also observe an aging effect which broads the metamagnetic-like transition suggesting the existence of Fe-Rh antiferromagnetic clusters which apparently are too small to be detected by the x-ray diffraction.The X-Ray diffractogram obtained in the Fe-Rh surface of the film do not show the Fe-Rh antiferromagnetic phase (a=2.98 $\AA$) but peaks at same locations of those found for the pure ordered Fe$_3$Pt film (a=3.72 $\AA$).  
\end{abstract}
\pacs{} 
\maketitle 
The Fe-Rh system in the composition $Fe_{1-x}Rh_x$ with x$\approx 0.5$ can crystalize under certain heat-treatment conditions in the CsCl structure \cite{k1,k2,k3} . In this cubic phase, equiatomic Fe-Rh is antiferromagnetic and exhibit a metamagnetic magnetostructural phase transition at ambient temperature upon applied magnetic fields of the order of Teslas \cite{ding}. The transition temperature is highly sensitivy to small changes out of the equiatomic stoichiometry \cite{kuba} and microstructural scale \cite{kang,marquina}. The first order magnetostructural phase transition presents a volume expansion of about 1$\%$ when entering in the ferromagnetic phase and a temperature hysteresis of the order of $\approx 10K$ \cite{k1,maat,ibarra,anna}. Films of this system has potential for applications as micro-electromechanical devices \cite{thiele}. In this work, we study $Fe_{1-x}Rh_x$ in the composition x$\approx 0.5$  obtained by electrodeposition on a thin foil of ordered Fe$_3$Pt. The work address the possibility of to obtain a thin film of Fe-Rh in the antiferromagnetic phase deposited on a ferromagnetic compound which would reduce the strength of the applied magnetic field necessary for the metamagnetic transition to occur at lower temperatures. We choose the ferromagnetic systems to be ordered Fe$_3$Pt which is a well know ferromagnet \cite{} for which the preparation of a thin foil is quite simple \cite{ercan}. The preparation and study of such a double-compound mostly motivated us. This work reports the first results obtained in the double-compound $Fe_3Pt-Fe-Rh$ prior to any heat treatment. As it is shown below, despite the Fe-Rh seems not to be in the antiferromagnetic state, the system $Fe_3Pt-Fe-Rh$ as prepared, presents an anomaly near 120 K in the temperature dependent magnetization curve when cooled from 300 K in low applied magnetic fields. The anomaly appears for fields as low as 20 Oe and it is reproducible resembling a metamagnetic transition. 

The $Fe_{1-x}Rh_x$ with x$\approx 0.5$ thin film was obtained by electrodeposition on a 0.2 $cm^2$ of an ordered $Fe_3Pt$ thin foil.  The electrodeposition occur in galvanostatic conditions from a bath composed of: 0.01 M $Fe_2(SO_4)_3$, 0.001- 0.0001 M $Rh_2(SO_4)_3$, 0.1 M $K_2SO_4$ and 0.05 M Sodium citrate ($Na_3C_6H_5O_7$). The final film thickness of Fe-Rh was about 30-50 nm depending on the applied current density  which was varied from 1 to 5 mA$cm^{-2}$ and the charge for preparing all deposits was kept to 10 C \cite{noce,taba,schulz}. An EG-G PAR potentiostat-galvanostat, model 273, served as a constant current source.  The electrodeposition occur under previously established conditions  producing an approximately equiatomic deposition of Fe and Rh \cite{noce}. The $Fe_3Pt$ substrate with $\approx$3 $\mu$m thickness was obtained by cold-rolling a previously prepared arc-melted sample. The ordered phase of the film was achieved after an appropriated heat treatment as described in Ref.\onlinecite{ercan}.  Figure 1a shows the X-ray diffractogram obtained for the $Fe_3Pt$ foil only, which is plotted with the diffractogram  of the Fe-Rh surface. It is possible to see in Fig. 1a that the diffraction peaks obtained from the Fe-Rh surface of the film are coincident with the $Fe_3Pt$ peaks (the arrows show the $Fe_3Pt$ peaks) suggesting that the final Fe-Rh thin film was deposited with the same lattice parameter of the substrate $Fe_3Pt$, a=3.72$\AA$. This fact suggests absence of the antiferromagnetic phase of equiatomic Fe-Rh, which has a lattice parameter  a=2.95$\AA$. 

Magnetisation data were obtained by using a PPMS-9T Quantum-Design magnetometer. All data were obtained with the magnetic field applied parallel to the plane of the films. Isofield curves, $M vs. T$, and isothermal curves, $M vs. H$,  were all obtained after cooling the sample from 300 K, in zero magnetic field (zfc), but in the presence of the earth magnetic field, to a desired temperature. After that, for isofield $M vs. T$ curves, a magnetic field was applied reaching the desired value without overshooting and data were obtained by heating the sample at fixed increments of temperature up to 300 K, which is followed by data acquisition cooling the sample at fixed $\delta T$ increments down to 1.8 K (10 K for H= 10 kOe curves) followed one more time by data acquisition heating the sample up to 300 K. These procedures correspond to a cycle starting with a zfc (zero-field-cooled) curve obtained by heating from 1.8 K followed by a fcc (field-cooled-cooling) curve  followed by a fch (field-cooled-heating) respectively. Isofield $Mvs.T$ curves where obtained for fields ranging from 20 Oe to 10 kOe. For isothermals $M vs. H$ curves, magnetisation is measured as the magnetic field increases (or decreases depending on the branch of the hysteresis curve) at fixed increments.

\begin{figure}
\includegraphics[width=\linewidth]{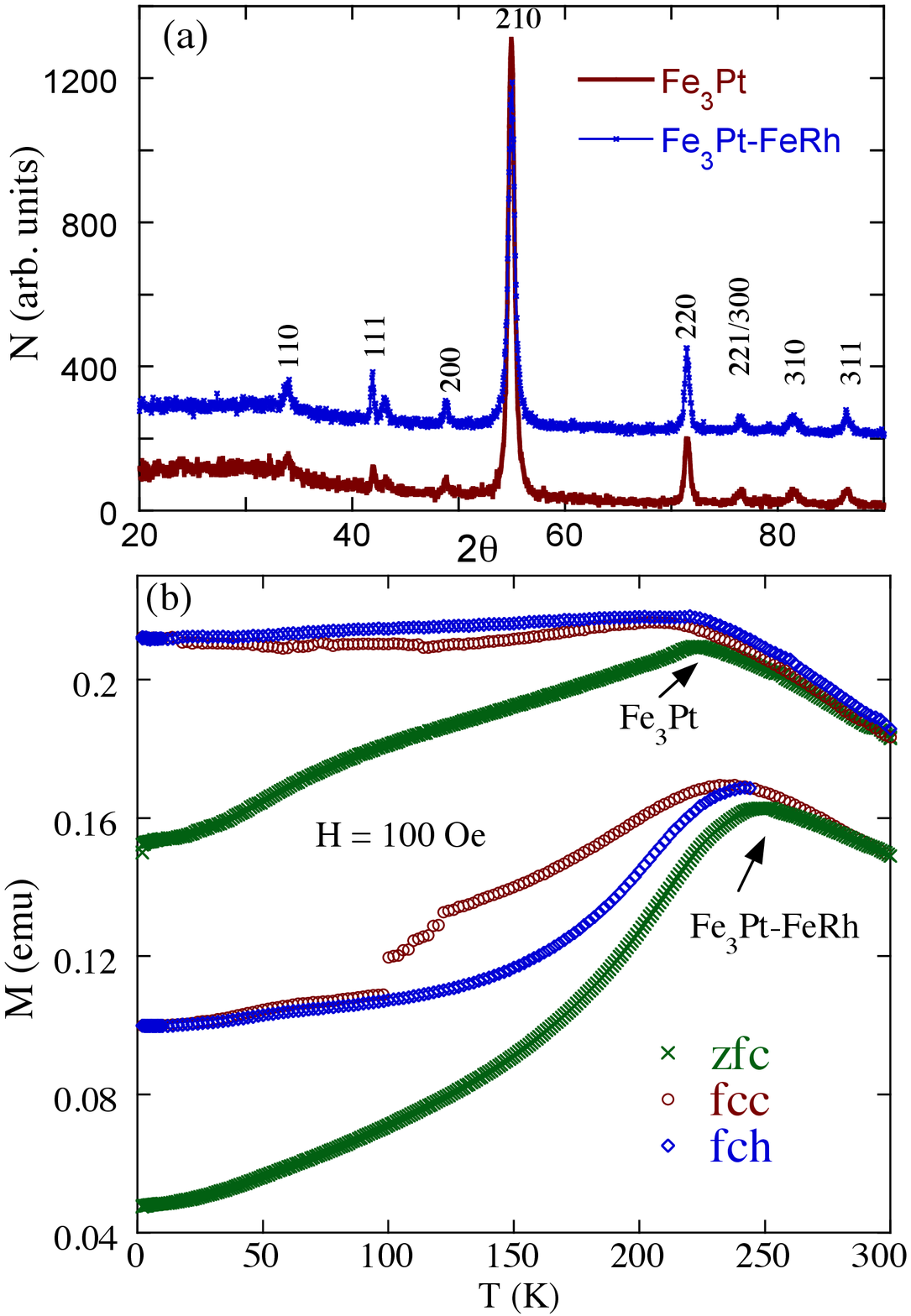}
\caption{\label{fig1}{a) X-ray diffractogram of $Fe_3Pt$ and $Fe_3Pt-FeRh$ films. b) Magnetisation curves for H = 100 Oe for $Fe_3Pt$ and $Fe_3Pt-FeRh$ films.}} 
\end{figure}

We plot in Fig.1b magnetisation $M vs. T$ curves obtained with a magnetic field H = 100 Oe applied on pure $Fe_3Pt$ thin foil and on the system  $Fe_3Pt-Fe_{1-x}Rh_x$. The curves of each system are shifted along the Y-axis for better presentation. Figure 1b allows to visualize the overall effect of the thin Fe-Rh layer deposited on $Fe_3Pt$ foil. It is possible to see in this figure, that the fcc and fch curves for $Fe_3Pt$ are much closer below 200 K than for the $Fe_3Pt-FeRh$ system and also, contrary to what occurs for $Fe_3Pt-FeRh$, the fch curve  for $Fe_3Pt$ lies above the fcc curve.  This figure also shows an anomaly in the fcc curve of the system $Fe_3Pt-FeRh$, discussed below, suggesting a kind of a metamagnetic transition.

\begin{figure}
\includegraphics[width=\linewidth]{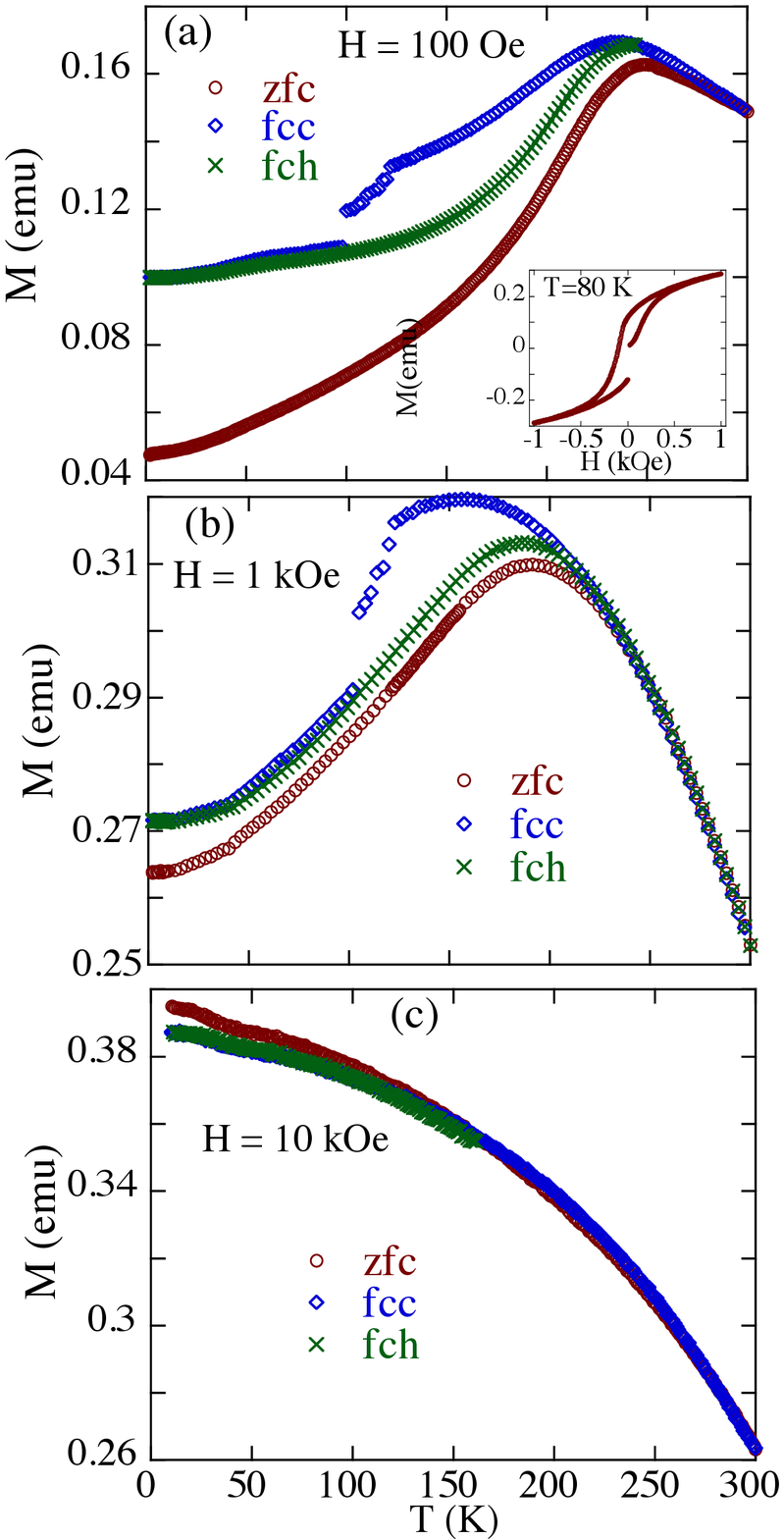}
\caption{\label{fig2}{Magnetization curves as obtained for the $Fe_3Pt-FeRh$ film for: a) 100 Oe; b) 1 kOe; c) 10 kOe. Inset a) M vs. H curve for T = 80 K}}
\end{figure}

Figure 2 shows isofield curves as obtained for $Fe_3Pt-Fe_{1-x}Rh_x$ sample with H=100 Oe (Fig.2a), 1 kOe (Fig.2b) and 10 kOe (Fig.2c) (Fig.2a is the same of Fig.1b and it is also presented in this figure for the sake of comparison). It is possible to see on Figs.2a and 2b, an anomalous effect occurring on the fcc curves, where magnetization starts dropping to lower values as temperature is cooling down below $\approx$120 K. The size, temperature position, and width of the  anomaly in the magnetisation fcc curves, which resembles a metamagnetic transition, do not show considerable changes with field, as Fig. 2b was obtained with H = 1 kOe and Fig. 2a with H = 100 Oe. For H = 10 kOe (Fig. 2c)  the anomalous effect is absent, and interestingly, the zfc curve appears above the fcc and fch curves (the fcc and fch curves for H=10 kOe are virtually coincident), while the zfc curves in Fig. 2a and 2b  appears below the fcc and fch curves. These differences are probably related to the alignment of ferromagnetic domains which depends on the strength of the magnetic field.  It is interesting to observe that at temperatures above 120 K, the fcc curves in Figs. 2a and 2b lies well above the zfc and fch curves, but these fcc curves lie almost perfectly over the fch curves below 100 K. These facts suggest the existence of a kind of an antiferromagnetic phase below 100 K, which apparently for the zfc and fch branches continuously evolve to a ferromagnetic phase as heating above 100 K. 
To check for this apparent antiferromagnetic phase below 100 K, we measure an isothermal $M vs. H$ curve at 80 K, depicted in the inset of Fig.2a, which show a typical ferromagnetic hysteresis curve. One may understand these facts as a coexistence of two phases below 100 K, a ferromagnetic phase due to the $Fe_3Pt$ and an antiferromagnetic phase due to the Fe-Rh thin layer, but this conclusion is in apparent contradiction with the X-ray results showing a typical ferromagnetic Fe-Rh lattice parameter. 

\begin{figure}
\includegraphics[width=\linewidth]{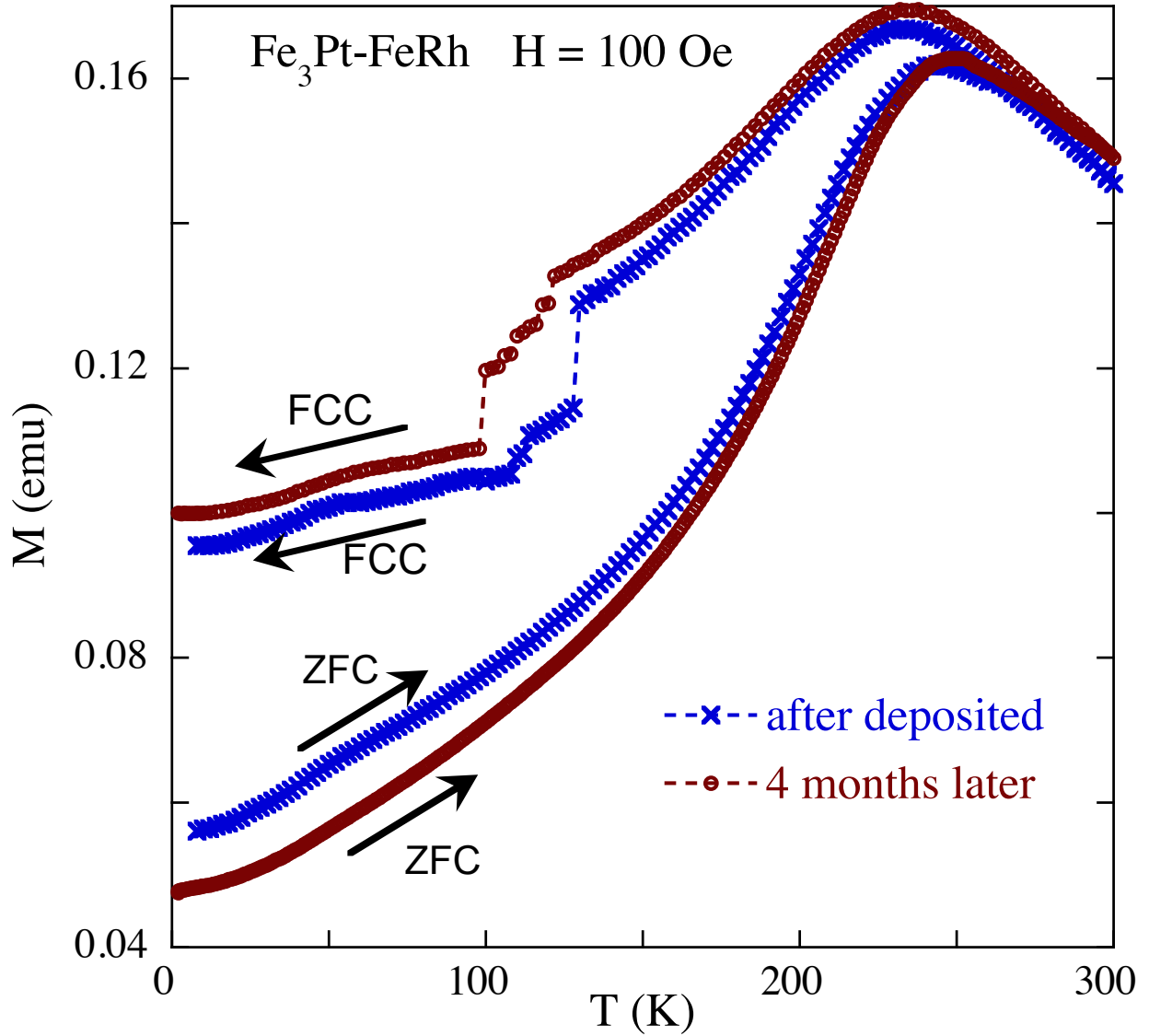}
\caption{\label{fig3}{Magnetization curves as obtained for the $Fe_3Pt-FeRh$ film for 100 Oe with 4 months delay.}} 
\end{figure} 

Figure 3 show two measurements obtained in the same sample at same conditions but with approximately 4 months delay between each other which show some kind of aging or time annealing effect which shift  the onset temperature of the anomaly and broads its width. One may observe that the curves presented in Fig.~\ref{fig2} correspond to data obtained $\approx$4 months after the eletrodeposition. We speculate that this aging effect might be due to small domains or clusters of  Fe-Rh in the antiferromagnetic phase (CsCl structure) which are too small to be detected by our X-ray diffraction technique (we lack on low-angle diffraction facilities), but large enough to produce the observed anomalous magnetisation on the fcc curves.

\begin{figure}
\includegraphics[width=\linewidth]{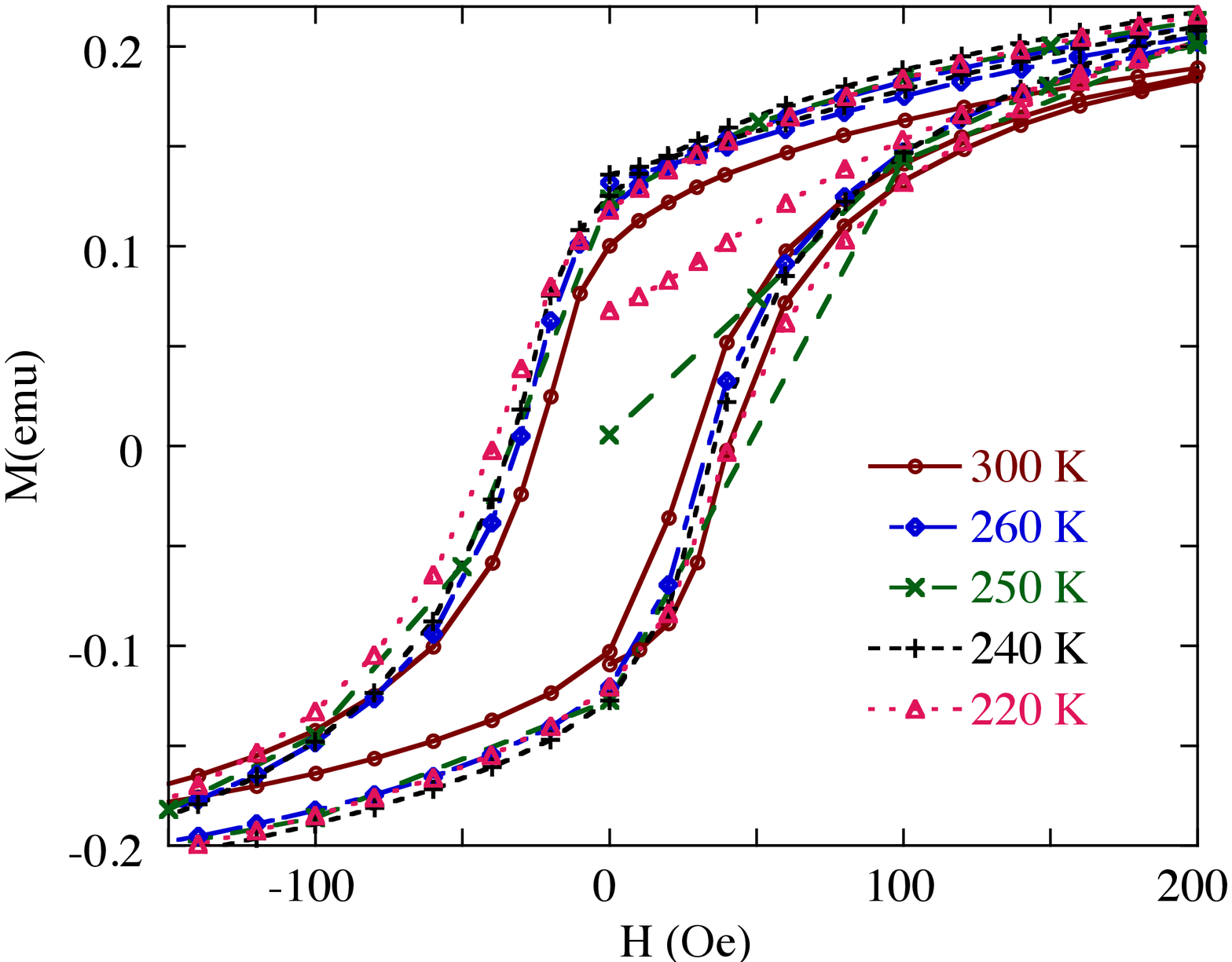}
\caption{\label{fig4}{isothermal M vs. H curves obtained for several temperatures.}} 
\end{figure} 

One should note that  the isofields $Mvs.T$ curves of Fig. 2 show a maximum in the magnetisation located approximately at 240-250 K. To check for possible effects related to this maximum on $Mvs.H$ curves  we obtained few isothermals for temperatures around 250 K which are presented in Fig.~\ref{fig4}. The curves in Fig.~\ref{fig4} show a typical ferromagnetic hysteresis behavior with no apparent change in the magnetisation isothermals as the maximum in the $Mvs.T$ curves at $\approx$ 250 K is crossed. 
We mention (not shown) that we check for a possible frequency dependence of the anomaly by measuring ac susceptibility curves for three different frequencies, 50, 500 and 2000 Hz with an ac magnetic field of 1 Oe and a dc magnetic field of 20 Oe. The resulting curves of the susceptibility amplitude versus temperature did not show any frequency dependent effect. We also mention that we search for time relaxation effects when in the fcc curve near the anomalous transition above 120 K, but no magnetic relaxation was observed.

In conclusion we obtained a $Fe_3Pt-FeRh$ system by electrodeposition of Fe-Rh  on a $Fe_3Pt$ ordered foil. The resulting $Fe_3Pt-FeRh$ system show, prior further heat-treatments, an anomaly in the magnetization when the sample is cooled in a magnetic field from 300K down to lower temperatures, after reaching 300 K with the same applied field from a zero-field-cooled heating cycle from 1.8 K .  The anomalous magnetization is reproducible and appears at temperatures close to 120 K from fields as low as 20 Oe up to fields as high as 1 kOe and resembles a metamagnetic transition.  The X-ray diffractogram of the film shows absence of the expected antiferromagnetic phase of equiatomic Fe-Rh which would explain such a kind of metamagnetic transition. We also observe that the anomalous magnetisation effect suffer an aging effect suggesting the existence of equiatomic Fe-Rh clusters in the antiferromagnetic phase, which are probably too small to be detected by the X-ray diffractogram.

SSS,ADA and FGG acknowledges support from CNPq, Brazilian Agency. 
 

\begin{thebibliography}{99}
 \bibitem{k1}J. S. Kouvel and C. C. Hartelius, J. Appl. Phys.{\bf 33}, 1343 (1962).
\bibitem{k2}J. S. Kouvel, J. Appl. Phys. {\bf 37},1257 (1966).
\bibitem{k3}J.M. Lommel and J.S. Kouvel, J. Appl. Phys. {\bf 38}, 1263 (1967). 
\bibitem{ding}Y. Ding, D.A. Arena, J. Dvorak, M. Ali, C. J. Kinane, C.H. Marrows, B.J. Hickey, and L.H. Lewis, J. Appl. Phys. {\bf 103}, 07B515 (2008). 
\bibitem{kuba}O. Kubaschewski, IRON-Binary Phase Diagrams, Springer, Berlin, 1982.
\bibitem{kang}K. Kang, A. R. Moodenbaugh, and L. H. Lewis, Appl. Phys. Lett. {\bf 90}, 153112  (2007).
\bibitem{marquina}C. Marquina, M.R. Ibarra, P.A. Algarabel, A. Hernando, P. Crespo, P. Agudo, and A.R. Yavari, and E. Navarro, J. Appl. Phys. {\bf 81}, 2315 (1997).
\bibitem{maat}S. Maat, J. U. Thiele, and E. E. Fullerton, Phys. Rev. B {\bf 72}, 214432 (2005).
\bibitem{ibarra}M. R. Ibarra and P. A. Algarabel, Phys. Rev. B {\bf 50}, 4196 (1994). 
\bibitem{anna}M. P. Annaorazov, S. A. Nikitin, A. L. Tyurin, K. A. Asatryan, and A. K. Dovletov, J. Appl. Phys. {\bf 79}, 1689 (1996).
\bibitem{thiele}J.-U. Thiele, S. Maat, and E. E. Fullerton, Appl. Phys. Lett. {\bf 82}, 2859 7 (2003).
\bibitem{ercan}E.E. Alp, M. Ramanathan, S. SalemSugui, F. Oliver, V. Stojanoff, and D.P. Siddons, Rev. Sci. Instrum. {\bf 63}, 1221 (1992).
\bibitem{noce}R.D. Noce, unpublished
\bibitem{taba}I. Tabakovic, S. Riemer, V. Vas'ko, and M. Kief, Electrochim. Acta {\bf 53}, 2483 (2008).
\bibitem{schulz}E.N. Schulz, D.R. Salinas, S.G. Garcia,  Electrochem. Commun. {\bf 12}, 583 (2010).
\end{thebibliography}
\end{document}